\documentclass[epjCONF]{svjour}
\usepackage{graphics}
\usepackage[varg]{txfonts} 
\usepackage[latin1]{inputenc}
\session-title{A Universe of Dwarf Galaxies: Observations, Theories, Simulations}
\begin{document}
\title{Detailed Chemical Abundances of Globular Clusters in Local Group Dwarf Galaxies}
\author{J. E. Colucci\inst{1}\fnmsep\thanks{\email{jcolucci@ucolick.org}} \and R. A. Bernstein\inst{1}  \and A. McWilliam\inst{2}}
\institute{Department of Astronomy and Astrophysics, 1156  High Street, UCO/Lick Observatory, University of California, Santa Cruz, CA 95064 \and The Observatories of the Carnegie Institute of Washington, 813 Santa Barbara Street, Pasadena, CA 91101-1292 }
\abstract{ We present detailed chemical abundances of Fe, Ca and Ba
  for 17 globular clusters (GCs) in 5 Local Group dwarf galaxies: NGC
  205, NGC 6822, WLM, the SMC and LMC.  These abundances are part of a
  larger sample of over 20 individual elements measured in GCs in
  these galaxies using a new analysis method for high resolution,
  integrated light spectra.  Our analysis also provides age and
  stellar population constraints.  The existence of GCs in dwarf
  galaxies with a range of ages implies that there were episodes of
  rapid star formation throughout the history of these galaxies; the
  abundance ratios of these clusters suggest that the duration of
  these burst varied considerably from galaxy to galaxy.  We find
  evolution of Fe, Ca, and Ba with age in the LMC, SMC, and NGC 6822
  that is consistent with extended, lower-efficiency SF between
  bursts, with an increasing contribution of low-metallicity AGB ejecta at late times.
  Our sample of GCs in NGC 205 and WLM are predominantly old and
  metal-poor with high [Ca/Fe] ratios, implying that the early history
  of these galaxies was marked by consistently high SF
  rates.} 

\maketitle
\section{Introduction}
\label{intro}
Stars of all ages record the chemical enrichment of a galaxy
throughout its history.  To unravel the complete formation history of
a galaxy, one must target both old and young stellar populations.
While very young stars are conveniently luminous, and detailed
abundances for a handful of stars have been obtained out to distances
of $\sim$900 kpc \cite{venn01}, older stars are far too faint for
detailed abundance analysis much beyond $\sim$100 kpc.  To enable
measurements in more distant galaxies, we have developed a new method
of abundance analysis that allows us to analyze high resolution
spectra of the integrated light (IL) of globular clusters (GCs), both
old and young.  The method has been developed using a ``training set''
of MW and LMC GCs and has been demonstrated to provide abundances as
accurate as obtained from individual stars --- $\pm$0.1 dex or better
for [Fe/H] and [X/Fe] \cite{mb08,c09,b10}.  Using this method, we can
target GC systems in galaxies at distances up to $\sim$4 Mpc using 8-m
class telescopes to probe their formation histories with the same
fidelity that stellar abundances have provided in the Milky Way.

In these proceedings, we present partial results for an ongoing survey
of detailed chemical abundances of GCs in Local Group dwarf galaxies.
Data were obtained for 17 GCs in NGC 205, NGC 6822, WLM, and the
Magellanic Clouds with the HIRES-r spectrograph on the Keck I
telescope, the MIKE spectrograph on the Magellan Clay telescope, and the echelle spectrograph on the  du Pont telescope.

\section{Results: Local Group Dwarf Galaxies }
\label{sec:1}
Our results to date show that these samples of GCs have ages and
abundances consistent with the field stars in their respective
galaxies (where such abundances are available), demonstrating that
these low mass galaxies had episodes of rapid star formation where
massive clusters were able to form out of the same gas reservoir as
field stars.  Results for [Fe/H], [Ca/Fe], and [Ba/Fe] for the 17 GCs
are listed in Table 1.  In Figure 1, we plot these along with
abundances taken from the literature for individual stars.

In the LMC, our cluster sample has [Ca/Fe] values that decrease with
both [Fe/H] and age.  This is consistent with available data for individual stars in the LMC
 \cite{mucc08,mucc10,pompeia08}, 
although few comparisons are available for the SMC.  Note that [Ca/Fe]
is representative of the true [$\alpha$/Fe] in the GC {\it integrated light}, whereas the
commonly-used low-resolution indexes that depend on [Mg/Fe] are
strongly affected by self-enrichment within the GC stars and
therefore not representative of the gas-abundance at the time of
formation (see \cite{c09}).  The old, metal-poor GC in NGC 6822 has
solar [Ca/Fe].  The 3 GCs sampled in NGC 205 and the sole GC in WLM
are reasonably metal-poor ([Fe/H]$<-1$), and are consistently enhanced
in $\alpha$-elements ([Ca/Fe]$>+0.2$).  We note that our findings for
NGC 205 are consistent with those of \cite{dacosta88}, who found these
GCs had properties similar to MW GCs using low resolution
spectroscopy.  Our results disagree, however, with the estimates of
solar [$\alpha$/Fe] by \cite{sharina06} based on low resolution indexes,  which we believe are unreliable, particularly at low metallicity (see \cite{c09}).

Ba, which is an s-process element, is  measured here for the first time in most of these
galaxies.  We find that it varies more in the dwarf galaxies than it does in the MW,
and is highest in young GCs in the LMC, SMC, and NGC 6822.  
High [Ba/Fe] in young GCs and the increasing [Ba/Fe] with decreasing age in the LMC is
consistent with  increasingly significant contributions from  low-metallicity AGB-stars at late times in these galaxies (see similar results found in \cite{pompeia08}).

While the samples in each galaxy are small, these abundances suggest
that the SF histories of Local Group dwarf galaxies are diverse; SF was extended  and occurred predominantly at low-efficiency in the LMC, SMC and NGC 6822 in between short bursts of more rapid SF that resulted in some massive, surviving GCs.  In NGC 205 and WLM, a larger fraction of the SF was rapid and efficient.

\section{Summary}
\label{sec:3}
We present the first detailed abundances for both old and young GCs
in NGC 205, NGC 6822, WLM, the LMC and the SMC
using a new analysis method.  Differences in the SF history between
normal galaxies (e.g. MW \& M31) and dwarf galaxies are clearly
evident in our high resolution analysis of GCs
\cite{mb08,c09,b10,c10}.

We now have the detailed abundances and ages necessary to begin to
quantify the chemical evolution of several dwarf galaxy GC systems.  Over 20 $\alpha-$, Fe-peak,
n-capture, and light element abundances, as well as age
measurements, will be presented in a future paper.  We are currently
obtaining data for the GC systems in several 
galaxies in and beyond the Local Group.

\begin{figure}
\resizebox{1.\columnwidth}{!}{
\includegraphics{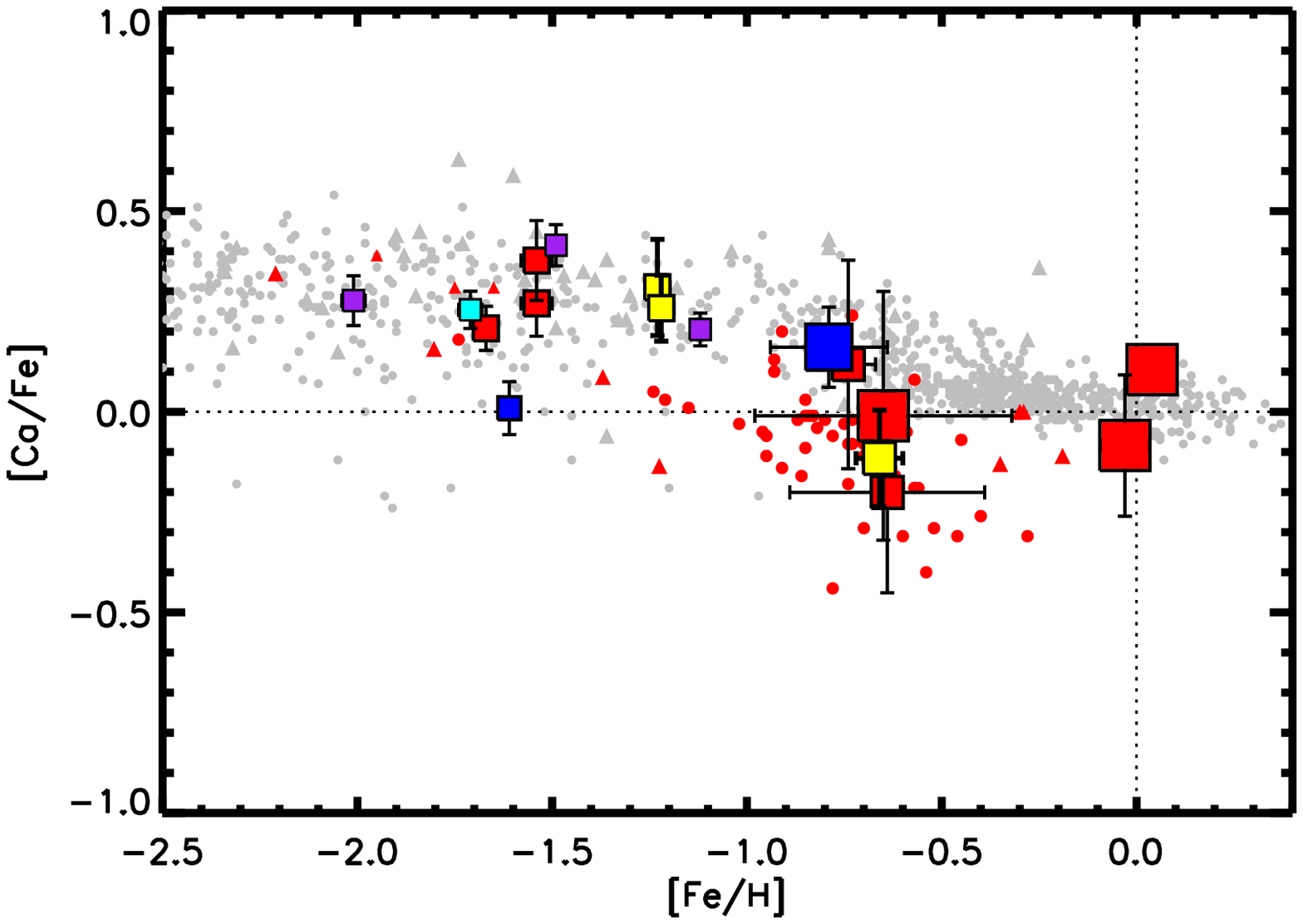}}
\resizebox{1.\columnwidth}{!}{
\includegraphics{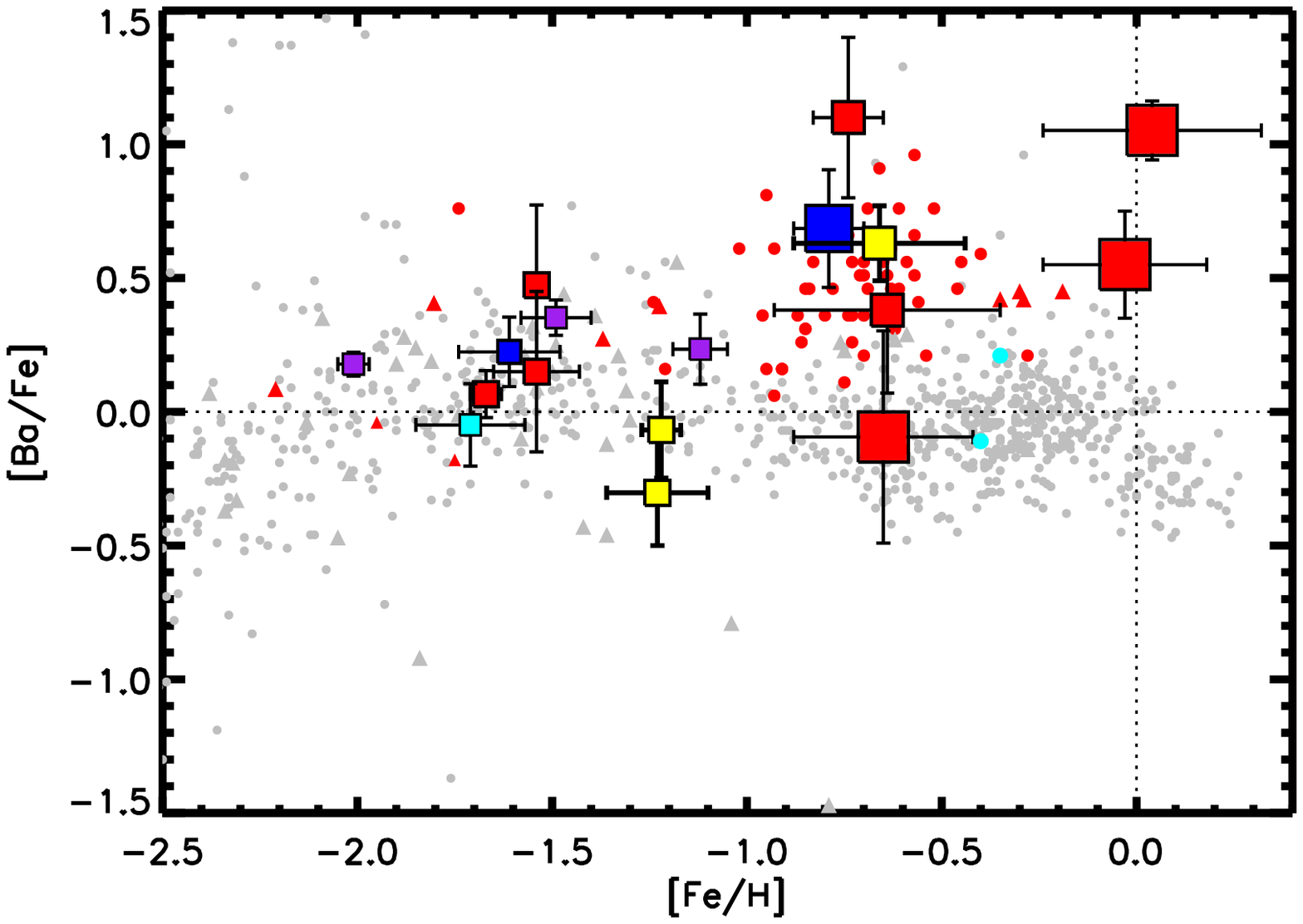}}
\caption{Abundance ratios for GCs in dwarf galaxies.  GCs in the LMC,
  SMC, NGC 205, WLM, and NGC 6822 are designated by red, yellow,
  purple, cyan, and blue squares, respectively. Small, medium, and large squares
  correspond  to GC age ranges of  $>$7 Gyr,1$\sim$6 Gyr, 0.01-1 Gy respectively.  Abundances from the
  literature \cite{venn01,mucc08,mucc10,venn04} for individual  stars in the MW, LMC, and WLM are shown by small gray,
  red and cyan points, respectively.}
\label{fig:1}       
\end{figure}


\begin{table}
{\footnotesize
\caption{ Abundances in  Local Group Dwarf Galaxies.}
\label{tab:1}       
\begin{tabular}{lccc}
\hline\noalign{\smallskip}
Name & [Fe/H] & [Ca/Fe] & [Ba/Fe] \\
\noalign{\smallskip}\hline\noalign{\smallskip}

\textbf{NGC 205}\\
  HI & $-1.49 \pm 0.02$ & $+0.42 \pm 0.05$ & $+0.35 \pm 0.07$  \\
  HII & $-1.12 \pm 0.02$ & $+0.21 \pm 0.04$ & $+0.23 \pm 0.13$\\
 HVIII & $-2.01 \pm 0.03$ & $+0.28 \pm 0.06$ & $+0.18 \pm 0.05$ \\
\noalign{\smallskip}\hline    
\textbf{WLM}\\
   WLM-GC& $-1.71 \pm 0.03$ & $+0.25\pm 0.05$ &$-0.05 \pm 0.15$\\
\noalign{\smallskip}\hline
\textbf{NGC 6822} \\
HVII& $-1.61 \pm 0.02$ & $+0.01\pm 0.07$ & $+0.22 \pm 0.13$\\
   HVI& $-0.79 \pm 0.15$ & $+0.16\pm 0.10$& $+0.69 \pm 0.22$ \\

\noalign{\smallskip}\hline
\textbf{SMC} \\
 NGC 121& $-1.23 \pm 0.03$ & $+0.31\pm 0.12$ & $-0.43 \pm 0.20$\\
NGC 416 & $-1.22 \pm 0.03$  & $+0.26 \pm 0.08$ & $-0.07 \pm 0.18$\\
 NGC 419 & $-0.66 \pm 0.06$  & $-0.12 \pm 0.12$ &$+0.63 \pm 0.14$\\
\noalign{\smallskip}\hline

\textbf{LMC} \\
   NGC 2005& $-1.54 \pm 0.04$ & $+0.27\pm 0.08$ &$+0.47 \pm 0.30$ \\
 NGC 2019& $-1.67 \pm 0.03$  & $+0.21 \pm 0.05$&$+0.07 \pm 0.09$ \\
 NGC 1916 & $-1.54 \pm 0.04$  & $-0.38 \pm 0.10$ & $+0.15 \pm 0.30$\\
   NGC 1978& $-0.74 \pm 0.07$ & $+0.12\pm 0.26$& $+1.10 \pm 0.09$ \\
 NGC 1718 & $-0.64 \pm 0.25$  & $+0.20 \pm 0.26$ & $+0.38 \pm 0.11$ \\
 NGC 1866 & $+0.04 \pm 0.04$  & $-0.10 \pm 0.04$ & $+1.05 \pm 0.13$\\
  NGC 1711& $-0.82 \pm 0.15$ & $-0.01 \pm 0.32$ &$-0.09 \pm 0.40$ \\
 NGC 2100 & \llap{$>$}$-0.03 \pm 0.06$  & $-0.08 \pm 0.18$& $+0.55 \pm 0.01$ \\

\noalign{\smallskip}\hline
\end{tabular}}
\end{table}

\end{document}